\documentclass[twocolumn,twocolappendix  ]{aastex631}
\usepackage{amsmath}

\shorttitle{SET-AI and PBH}
\shortauthors{Baghram}

\begin{document}
	
\title{In Search of Extraterrestrial Artificial Intelligence Through Dyson Sphere\textendash like structures around Primordial Black Holes\footnote{This work is dedicated to Sohrab Rahvar, whom we admire for first teaching us about life in the Universe.}}

\author[0000-0001-6131-4167]{Shant \surname{Baghram}}
\email{baghram@sharif.edu}
\affiliation{Department of Physics, Sharif University of Technology, Tehran 11155-9161, Iran}
\affiliation{Research Center for High Energy Physics, Department of Physics, Sharif University of Technology, Tehran 11155-9161, Iran}


\begin{abstract}
Are we alone? It is a compelling question that human beings have confronted for centuries. The search for extraterrestrial life is a broad range of quests for finding simple forms of life up to intelligent beings in the Universe. The plausible assumption is that there is a chance that intelligent life will be followed by advanced civilization equipped or even dominated by artificial intelligence (AI). In this work, we categorize advanced civilizations (on an equal footing, an AI-dominated civilization) on the Kardashev scale. We propose a new scale known as the space exploration distance to measure civilization advancement. We propose a relation between this length and the Kardashev scale. Then, we suggest the idea that advanced civilizations will use primordial black holes (PBHs) as sources of harvesting energy. We calculate the energy harvested by calculating the space exploration distance. Finally, we propose an observational method to detect the possibility of extraterrestrial AI using Dyson sphere\textendash like structures around PBHs in the Milky Way and other galaxies. 
\end{abstract}

\section{Introduction}
Are we alone? It is a compelling question that human beings have confronted for centuries from antiquity to Giordano Bruno to modern times with the quest for finding extraterrestrial life by the "search for extraterrestrial intelligence" project\footnote{See \url{https://www.seti.org/}}. Nowadays, this question has become part of scientific investigation in a discipline known as astrobiology. Searching for life in the Universe is a multidisciplinary inquiry rooted in biology and astronomy \citep{Plaxco:2001}. The expedition to find life can range from simple biological living systems to extraterrestrial intelligent life.
Accordingly, searching for life is a vast ground of research starting from finding extrasolar planets (hereafter exoplanets) in habitable zones \citep{2018Galax...6...51L} to finding radio signals from intelligent life \citep{Tarter:2001kq,2016ApJ...828...19R}.
It is almost three decades from the breakthrough discovery of the first exoplanet \citep{Mayor:1995}. Nowadays, we have found more than 5600 exoplanets \footnote{For the number of discovered and confirmed exoplanets, see \url{https://science.nasa.gov/exoplanets/}}.
The search for any biosignature from exoplanets is an extensive field of research. For this type of study see \cite{2005AsBio...5..706S, 2010ARA&A..48..631S,2018AsBio..18..663S} and references therein. \\
Parallel to these studies, the question of how extraterrestrial life could be intellectually complicated and intelligent becomes a burning question. In this direction, the Drake equation is a step forward to quantify the probability of finding intelligent life in the Universe \citep{Drake1961}. Three years after Drake's insightful idea, Nikolai Kardashev, a soviet astronomer, attended an astronomy conference at Byurakan Observatory, Armenia, in 1964. At this meeting, Kardashev categorized the three levels of civilization based on their consumption of energy \citep{Kardashev1964}. The first category is those who consume all the energy available to the planet. The second category uses all the energy of the host star by a method like a Dyson sphere \footnote{{{The Dyson sphere is a hypothetical space-based artificial megastructure around a star or other celestial energy source, which is used to collect effectively the energy for an advanced civilization. The first modern imagination of this structure is from a science fiction novel, "Star Maker" by  Olaf Stapledon in 1937 \cite{{Dyson:2017}}. It is realized by a thought experiment presented by Freeman Dyson in his 1960 paper  "Search for Artificial Stellar Sources of Infrared Radiation"\citep{Dyson1960a, Dyson1960b}}.}} \citep{Dyson1960a, Dyson1960b}, and the third level of civilization is related to the ones that use all the energy of the host galaxy. \\ 
{{In this sequence, we are confronted with more advanced civilizations. In our case, on Earth, as an example of a planet which hosts intelligent life, it seems that after a century of the invention of Guglielmo Marconi, our species is on the verge of using artificial intelligence (AI) and the rise of AI is on the horizon.
Due to recent developments, there is a nonzero chance that AI will overtake and dominate the planet \citep{Tegmark:2017}. In other words, advanced civilizations become defined on the same footing as civilizations based on or dominated by AI \footnote{The result of this study is also validated if you replace the AI with an advanced civilization}. 
Accordingly, the next question will be: what is the probability of finding extraterrestrial artificial intelligence (ET-AI) in the Universe? How we can categorize it using the Kardashev scale. And what is our plan for the search for extraterrestrial artificial intelligence (SET-AI)? }}  \\ One of the main characteristics of AI is its capability for calculating and storing data. It means that energy consumption is still a priority for our descendants. Also, there is a plausible chance that they use quantum computers as calculating machines \citep{Cherrat2022}. \\
In this work, we introduce a new length scale as the {\it{space exploration distance}} (SED), which is in a tight relation with the Kardashev scale. We assert that ET-AI will be on the Kardeshev scale between two and three. Accordingly, explore is explored further from the host planet and star. Then, we estimate the energy harvested from the sources besides the host star. These extra sources could be the stellar or/and primordial black holes (PBH). PBHs are one of the compelling candidates for dark matter (DM) \citep{Green:2020jor} and the most abundant and reliable energy source.\\ 
The structure of the work is as follows: In section \ref{Sec2}, we introduce the theoretical background of this work. In section \ref{Sec3}, we present our results on SET-AI, and in the final section \ref{Sec4}, we conclude. 
\section{Theoretical background}\label{Sec2}
In this section, we review the Drake equation, the idea of using a Dyson sphere for harnessing energy, and PBHs as a plausible candidate of DM in three upcoming subsections.
\subsection{Modified Drake equation for Extraterrestrial Artificial Intelligence}
The Drake equation encapsulates the probability of finding an intelligent civilization that desires to communicate  \citep{Drake1961}. The communication is assumed to be in a radio signal, and it is formulated as
\begin{equation}
	N_c=R_* \times f_{gs} \times f_p \times n_e \times f_l \times f_i \times f_c \times L_c,	
\end{equation}
where $R_*$ is the rate of star formation in solar masses per year, $f_{gs}$ is the fraction of the good stars, $f_p$ is the fraction of the stars that form planets, $n_e$ is the number of Earth-like planets, $f_l$ is the fraction of the planets where life originates, $f_i$ is the fraction planets that harbor intelligent life, $f_c$ is the fraction of planets that have intelligent life which has and desires to have interstellar communication, and finally $L_c$ is the lifetime of the communicative intelligent species.\\
Now, for finding an ET-AI, even beyond our Galaxy, using the idea of space megastructures like Dyson spheres, we modify the Drake equation as below
\begin{equation} \label{eq:Drake-Modified}
	N_{\rm{\tiny{ET-AI}}} =  N_g \times f_{gg}\times R_* \times f_{gs} \times f_p \times n_e \times f_l \times f_i \times f_{\rm{AI}} \times L_{\rm{AI}},	
\end{equation}
where $N_g$ is the number of galaxies, which we define shortly. $f_{gg}$ is the fraction of good galaxies that can harbor life.
Note that the star formation rate is a function of stellar mass and, in consequence, a function of the galaxy's total mass \citep{2022A&A...666A.186D}.
$f_{\rm{AI}}$ is the fraction of planets that host intelligent life, and their dominant descendants are  AI.  $L_{\rm{AI}}$ is the lifespan of a typical AI-dominated civilization.
As one may notice, we omit the probability of communicative intelligence fraction from the modified Drake equation \ref{eq:Drake-Modified}. This is done because our search method for ET-AI is not based on radio communication, and probably ET-AI are not social!
We should note that two unknowns of $f_{gs}$ and $f_{gg}$ representing "good stars" and "good galaxies" are themselves a topic of ongoing research (see \cite{2015ApJ...810L...2D,2020MNRAS.494.3048W,2021ApJ...919....8L}). 
{{It is worth mentioning that most of the galaxies, due to their numbers of stars, are likely to harbor life, and we can set $f_{gg}\simeq 1$. However, the work by \cite{2015ApJ...810L...2D} is an example of a study to answer the question "which type of galaxy is most likely to host complex life in the cosmos?"}}
Coming back to the number of galaxies $N_g$, one knows that the preparation of the data and using volume-limited constraints is a complicated survey-dependent procedure which reduces the number of observed galaxies in a specific redshift span \citep{2001AJ....122.1104Y}.
To obtain the number of galaxies $N_g$, which is calculated in the mass range of $(M_i,M_f)$ and redshift bin of $(z_i,z_f)$ we define
\begin{equation}
	N_g =	\int_{z_i}^{z_f} \frac{cdz}{H(z)}D^2_A \int_{\Delta\Omega}  d\Omega  \int_{M_i}^{M_f} \frac{dM}{M}  \frac{dn}{d\ln M}(M,z),
\end{equation}
where $H(z)$ is the Hubble parameter, $D_A$ is the comoving angular diameter distance, and $\Delta\Omega$ is the solid angle of a specific survey. $\frac{dn}{d\ln M}(M,z)$ is the comoving number density of DM halos \citep{1974ApJ...187..425P,2002PhR...372....1C} which host a central galaxy with the characteristics suitable for Drake equation. 
We should also note that at higher redshifts, the probability of finding AI/advanced civilizations should be smaller due to the lack of time for the development of civilization. Additionally, as mentioned, the probability decreases because galaxies appear fainter and smaller at higher redshifts (both intrinsically and observationally).
In conclusion, the modified Drake equation sets the arena of SET-AI beyond our Milky Way.
\subsection{ Dyson spheres}

An AI-based civilization,  as a continuation of intelligent life, will be distinguished from its ancestors by the capability of using energy effectively and vastly. We assume that the ET-AI, at least, is in a second-level civilization on the Kardashev scale. ET-AI uses space megastructures such as Dyson spheres for harnessing the energy of their host star \citep{Dyson1960a,Dyson1960b} and beyond. On the other hand, an AI-based society is computational with processing systems, which need a low temperature for high performance and, consequently, an enormous amount of energy.
Accordingly, they will make space constructions like Dyson spheres at a distance from the host star or black hole to use all energy of the source at the desired temperature.
These structures are heated up and will radiate in infrared \citep{2014ApJ...792...26W}. This idea motivates different groups to study excess infrared radiation in the spectrum of stars as potential Dyson spheres \citep{2024AJ....168..157C,2024MNRAS.531..695S}. 
On the other hand, they will use the shielding structures to bring the computational side of the structure to the temperature near interstellar medium in a typical galaxy.

The luminosity of a typical host star like the sun on the main sequence with a radius of $R_{\odot}\sim 7\times 10^8$ m is about $L_{\odot}\sim 3.8\times 10^{26}$ W.
It is a good approximation that a Dyson sphere acts as a blackbody. Accordingly, the luminosity per unit area $j$ is related  to temperature via Stephan-Boltzmann law:
\begin{equation}
	j=\sigma T^4,
\end{equation}
where $T$ is the temperature of the blackbody in kelvins and the Stefan–Boltzmann constant is $\sigma \simeq 5.7\times 10^{-8}  ~{\text{W} / m^2 {}\text{K}}^4$. So, the total power of the sun can be related to its temperature ($T_{\odot}\sim 5770$ K) as
\begin{equation}
	P_{\odot}=4\pi R^2_{\odot}\sigma T_{\odot}^4 \simeq 3.8\times 10^{26} {\rm{W}}.
\end{equation}
More advanced civilizations start to explore the space beyond their stellar neighborhood and harness the energy of nearby stars.
In this case, the typical radius of the Dyson sphere around a typical star $R_{\rm{Dyson}}|_{\rm{star}}$ can be approximated
\begin{equation}
	R_{\rm{Dyson}}|_{\rm{star}}\sim R_{\odot} (\frac{T_{\odot}}{T_{\rm{Dyson}}})^2,
\end{equation}
where $T_{\rm{Dyson}}$ is the blackbody temperature of the Dyson sphere. If we set the temperature of the Dyson sphere to $\sim 3000 ~{\rm{K}}$ to avoid realistic solid material from melting, Dyson sphere\textendash like space megastructures will be on the scale of $R_{\rm{Dyson}}|_{\rm{star}} \simeq 1.7 \times 10^{-2} \rm{AU}$. For a temperature of $300~ {\rm{K}}$ for having liquid water or a classical computing regime the scale is $R_{\rm{Dyson}}|_{\rm{star}} \simeq 1.7  \rm{AU}$ and for low-temperature computation around $30 {\rm{K}}$  the scale will $R_{\rm{Dyson}}|_{\rm{star}} \simeq 170  \rm{AU}$. \\
This range of temperatures suggests that we can find  Dyson sphere\textendash like structures as excess radiation in the peak wavelength obtained from Wien's law $\lambda_{\rm{max}} = \frac{b}{T}$,
where $T$ is the absolute temperature and $b\simeq 2.898 \times 10^{-3} {\rm{mK}}$ is a constant of proportionality called Wien's displacement constant.

\cite{Hsiao:2021qij} shows that BHs can be a more effective and promising energy sources than the main-sequence stars.
The major and main contribution from a BH can be obtained from accretion disks. The matter in the accretion disk is heated up by friction due to the strong gravity of the BH.
For a BH with mass $M$, the Eddington limit is \citep{Hsiao:2021qij}
\begin{equation}
	L_{\rm{Edd}} = \frac{4\pi GMm_p c}{\sigma_T}\sim 3.2 \times 10^4 (\frac{M}{M_{\odot}})L_{\odot},
\end{equation}
where $m_p$ is the mass of the proton and $\sigma_T=6.65\times 10^{-29}~ {\rm{m}}^{-2}$ is the Thomson scattering cross section.
The luminosity of the disk with the efficiency of $\eta_{\rm{disk}}$ and the accretion rate of $dm/dt$ is
\begin{equation}
	L_{\rm{disk}}=\eta_{\rm{disk}}\frac{dm}{dt}c^2,
\end{equation}
{{where for an extreme Kerr BH we have $\eta_{\rm{disk}}=0.399$ and for a nonrotating BH $\eta_{\rm{disk}}=0.057$ \citep{1974ApJ...191..507T}. $L_{\rm{disk}}$ depends on the accretion rate, which can be in the range of $10^{-15}-10^{-10} M_{\odot}s^{-1}$. For simplicity of calculation, we assume a PBH with a mass range of  $0.01-100 M_{\odot}$ with a conservative assumption of $\eta_{\rm{disk}}\sim 10^{-4}$, and we set $L_{\rm{disk}} = 10^{-4} 	L_{\rm{Edd}}$ which introduce powers of the order of $0.032-320 L_{\odot}$.
BHs are categorized into families of stellar BHs, intermediate-mass BHs, supermassive BHs.
		In the next subsection, we will briefly introduce PBHs as a candidates for DM. As we have not observed PBHs yet, one would question the assumptions, especially regarding the luminosity of their accretion disks and their mass, redshift, and environment dependence. However, the assumption of an accretion disk around PBHs is discussed in various works, see \cite{2008ApJ...680..829R, 2017PhRvD..95d3534A, 2018MNRAS.479..315S,2020A&A...642L...6B} and references therein. As mentioned in our proposal, we choose a conservative accretion efficiency. Also, we assume that this efficiency is mass, redshift, and environment independent.  }}

Around a PBH, one can make Dyson sphere\textendash like structures closer due to the small size of the accretion disk, approximately in the distance of 
\begin{equation} \label{eq:Rdyson-pbh}
	R_{\rm{Dyson}}|_{\rm{\tiny{PBH}}}\sim 2.7 \times 10^7 {\eta^{\frac{1}{2}}_{\rm{disk}}} (\frac{M}{M_{\odot}})^{\frac{1}{2}} (\frac{\rm{Kelvin}}{T})^2  \rm{AU},	
\end{equation}

\begin{table}
	\begin{center}
		\begin{tabular}{ c | c | c |c }
			Dyson sphere Temp. & $R_{\rm{Dyson}}|_{\rm{star}}$& $R_{\rm{Dyson}}|_{\rm{\tiny{PBH}}}$ & $\lambda_{\rm{max}}$\\ 
			\hline \hline
			3000 K & $\sim0.017$ AU  &  $\sim0.03$ AU & $\sim 966 ~ {\rm{nm}}$  \\ 
			300 K & $\sim1.7$ AU  & $\sim 3.0$ AU  & $\sim 9.66 ~ \mu {\rm{m}}$ \\
			30 K & $\sim170$ AU  &  $\sim 300$ AU  & $\sim 96.6 ~\mu {\rm{m}}$\\ 
		\end{tabular}
		\caption{A table showing the Dyson sphere\textendash like structures' temperature (first column) and their distances from the energy source (star and Primordial Black Hole) in the second and third columns. The fourth column shows the peak of the wavelength in blackbody radiation from a Dyson sphere\textendash like structure. $R_{\rm{Dyson}}|_{\rm{\tiny{PBH}}}$ is calculated for a solar mass PBH with $\eta_{\rm{disk}} = 10^{-4}$.}
		\label{Table1}
	\end{center}
\end{table}
In Table \ref{Table1}, we list the distances of Dyson sphere\textendash like structures from the star or PBH corresponding to the desired temperature. In the final subsection, we discussed PBHs.
\subsection{Primordial black holes as dark matter}
Gravitational wave detections by LIGO \citep{2016PhRvL.116f1102A} again raise interest in PBHs \citep{2016PhRvL.116t1301B,2020ARNPS..70..355C}. PBHs which were introduced by Zel'dovich and Novikov \citep{1966AZh....43..758Z} are a by-product of the evolution of the Universe, which can be produced in the early Universe \citep{2022arXiv221105767E}.
However, there are many sophisticated observational constraints on the abundance of PBHs in different ranges of their mass \citep{2017PhRvD..96b3514C}.
In this work, we assume that a nonzero fraction of DM ($f_{\rm{PBH}}$) is PBHs. \\
In the realm of the standard model of cosmology and structure formation, galaxies reside in DM halos \citep{2010gfe..book.....M}. The distribution of DM halos nearly follows a Press-Schechter distribution consistent with hierarchical structure formation \citep{1974ApJ...187..425P}.  $N$-body simulations and observations of DM tracers on small scales show that the distribution of DM in halos is almost Universal. To be more precise, the DM halo density profile nearly follows the one proposed by Navarro–Frenk–White (NFW),  $\rho_{\text{\tiny{NFW}}}$ \citep{Navarro:1995iw}:
\begin{equation}
	\rho_{\text{\tiny{NFW}}}(r)=\frac{\rho_s}{\frac{r}{r_s} (1+\frac{r}{r_s})^2},
\end{equation}
where $\rho_s$ and $r_s$ are, respectively, the specific density and radius corresponding to each DM halo. The edge of each DM halo can be specified by its virial radius $r_{\rm{vir}}$, which can be related to the specific radius of the halo with concentration parameter $c$ as $r_{\rm{vir}}=cr_s$. The concentration parameter ranges from $4-40$ for different DM halos. The total mass of the DM halo $M_{\rm{halo}}$, also related to the specific radius and concentration as
\begin{equation}
	M_{\rm{halo}}=\int_0^{r_{\rm{vir}}} 4\pi r^2\rho_{\text{\tiny{NFW}}}(r)dr = 4\pi\rho_sr_s^3A_{\rm{NFW}},
\end{equation} 
where $A_{\rm{NFW}}=\left[\ln(1+c)-\frac{c}{1+c}\right]$. If we define the mean density of DM halo $\bar{\rho}_{\rm{halo}} = 3M_{\rm{halo}} / (4\pi r^3_{\rm{vir}})$, and the normalized distance to virial radius $x=r/r_{\rm{vir}}$, the NFW profile can be written as
\begin{equation}
	\rho_{\text{\tiny{NFW}}}(x)=\frac{\bar{\rho}_{\rm{halo}}}{3A_{\rm{NFW}}x(c^{-1}+x)^2}.
\end{equation}
As the distribution of PBHs in a DM halo is almost uniform, the NFW distribution is sufficient to obtain the number density of PBHs via the knowledge of $f_{\rm{PBH}}$ and the total mass of PBHs \footnote{In this work, we assume a monochromatic mass spectrum for PBHs \citep{Erfani:2021rmw,Kleban:2023ugf}.}.
In the next section, we will show the correspondence of civilization type and the number of PBH used for harvesting the energy by ET-AI.

\section{Search for Extraterrestrial Artificial Intelligence} \label{Sec3}
In this section, we realize the idea of using PBHs in the DM halo of a galaxy as a source of energy for an AI-based civilization. 
A type I civilization uses the energy of its host star that is incident on the planet with an energy scale of $\sim 4\times 10^{16}$ W. A type II civilization uses the whole energy of the host star, $\sim 4\times 10^{26}$ W and a type III uses whole energy of the galaxy, $\sim4\times 10^{37} $ W. For these scales, we assume a typical star like the Sun in a typical galaxy such as the Milky Way. Sagan introduced a civilization parameter due to the amount of energy consumption as \citep{Sagan2000}
\begin{equation}
	K \simeq 0.1 \left(\log (\frac{P}{1\rm{W}})-6\right),
\end{equation}
where $P$ is the consumed power.
In this conclusive section, we are in the position to search for ET-AI. We assume that they are living in a host planet-star system. However, they use energy storage from beyond their solar system for computation. Accordingly, we quantify the distance in which AI explores the space to collect the needed energy. \\
{{Now we define a specific length known as SED ${\cal{D}}_{\rm{sed}}$. We, human beings, as a civilization of $K \simeq 0.7$ \citep{1973Icar...19..350S}, install our facilities in the Lagrangian 2 point as a stable space construction, which exchanges the data. By stable construction, we mean structures that are in almost the same position with respect to the orbital position of the sun and Earth, like space telescopes at the Lagrangian 2 point. Besides them, we have already sent Voyager 1 and 2 almost beyond the solar system. They are still functional and collecting data \footnote{For more information, see \url{https://science.nasa.gov/mission/voyager} }. }}  Inspired by this observation and an extrapolation,  a type I civilization's structure will reach astronomical unit scales. So, type II civilizations will reach the gravitational boundaries of their solar system. We propose the approximate relation below between SED ${\cal{D}}_{\rm{sed}}$ and Sagan's type of civilization

\begin{table}
	\begin{center}
		\begin{tabular}{ c | c | c  }
			Civilization Type & Energy Cons. & SED: ${\cal{D}}_{\rm{sed}}$\\ 
			\hline \hline
			Our Civilization & $1.7\times 10^{13}$ W & $\sim0.01$ AU \\ 
			Type I & $4\times 10^{16}$ W & $\sim1-2$ AU \\
			Type II & $4\times 10^{26}$ W  & $\sim10^5$ AU\\ 
			Type III &$4\times 10^{37}$ W  & $\sim 4\times 10^{9}$ AU \\ 
		\end{tabular}
		\caption{Civilization Type in the first column is categorized vs. the energy consumption in the second column and Space Exploration distance ${\cal{D}}_{\rm{sed}}$ in the third column.}
		\label{Table2}
	\end{center}
\end{table}

\begin{equation}
	K \simeq 0.2\left(\log(\frac{{\cal{D}}_{\rm{sed}}}{1\rm{AU}}) +5\right). \label{eq:sed}
\end{equation}
To elaborate more, the equation \ref{eq:sed} shows that for our current civilization stage the ${\cal{D}}_{\rm{sed}} (K\simeq 0.7) \simeq 0.01 \rm{AU}$. {{As mentioned, we set the SED to the Sun-Earth Lagrangian 2 point, where we set our astronomical instruments (space structures). Nowadays the European Space Agency's (ESA) Gaia probe \citep{2016A&A...595A...1G}; the joint Russian\textendash  German high-energy astrophysics observatory Spektr\textendash RG, \footnote{official webpage: \url{https://www.srg.cosmos.ru/}} which carries both the extended Roentgen Survey with the Imaging Telescope Array (eROSITA) \citep{2012arXiv1209.3114M} and Astronomical Roentgen Telescope X-ray Concentrator (ART-XC) \citep{2021A&A...650A..42P}; the joint National Aeronautics and Space Administration, ESA and Canadian Space Agency (CSA) James Webb Space Telescope (JWST) \citep{2006SSRv..123..485G} and the ESA's Euclid mission \citep{2011arXiv1110.3193L}, are all at the Lagrangian 2 point. }} \\For a type I civilization, where all of the energy emitted from the planet is harvested, we use the astronomical unit scale. For a type II civilization, where all of the emitted energy of the star is harvested, the scale is up to the gravitational scale of the solar system, which is the Oort cloud. For the final stage, we set the scale of the extension of the disk of the Milky Way as a typical galaxy (see Table \ref{Table2}). 
This approximation can relate the amount of power consumed by a civilization to its SED with the relation 
\begin{equation}
	P \sim L_{\odot} \times \left(\frac{{\cal{D}}_{\rm{sed}}}{1\rm{pc}}\right)^2.
\end{equation}
Now, we want to calculate the number of Dyson sphere\textendash like structures as a function of the type of civilization. 
For this task, for a proof of concept, we assume a typical  Milky Way galaxy with a total mass of $M\simeq 1.5\times 10^{12} M_{\odot}$, the concentration of the dark matter halo profile $c \simeq 10$ and virial radius $r_{\rm{vir}}\simeq 200 ~\rm{kpc}$ \citep{2019MNRAS.487L..72G}. The fraction of PBHs $f_{\rm{PBH}}$ and their mass are free parameters.

The number of PBH around the radius of the SED is obtained as
\begin{equation}
	N_{\text{{PBH}}}(<{\cal{D}}_{\rm{sed}}) = \frac{f_{\text{\tiny{PBH}}}}{M_{\text{\tiny{PBH}}}}\int_{r_*}^{r_*+{\cal{D}}_{\rm{sed}}} \rho_{\text{\tiny{NFW}}}(r)4\pi r^2 dr,
\end{equation}
where $r_* \simeq 8 ~{\rm{kpc}}$ is the position of the planet in Milky Way. Here, we assume that the life-harboring planet whose inhabitants are AI resides far away from the center of the galaxy. However, as we know, the distribution in DM halo is spherical with a good approximation. The number of PBHs is
\begin{equation}
	N_{\rm{PBH}}(<{\cal{D}}_{\rm{sed}}) = f_{\text{\tiny{PBH}}}\frac{M_{\rm{halo}}}{M_{\text{\tiny{PBH}}}}{A^{-1}_{\rm{NFW}}} \times {\cal{F}}(r_*, {\cal{D}}_{\rm{sed}};c), \label{eq:NPBH}
\end{equation}	
where ${\cal{F}}(r_*, {\cal{D}}_{\rm{sed}};c)$ is a function of the concentration of DM halo and distance of the planet from the center of the galaxy and SED as below
\begin{equation}
	{\cal{F}}(r_*, {\cal{D}}_{\rm{sed}};c) = \ln(1+\frac{cD_*}{1+cx_*}) - \frac{cD_*}{(1+cx_*)(1+cx_*+cD_*)}, \label{eq:Fcal}
\end{equation}
where $x_*=r_* / r_{\rm{vir}}$ and $D_*= {\cal{D}}_{\rm{sed}} / r_{\rm{vir}}$.\\

\begin{figure}
	\centering
	\includegraphics[scale=0.55]{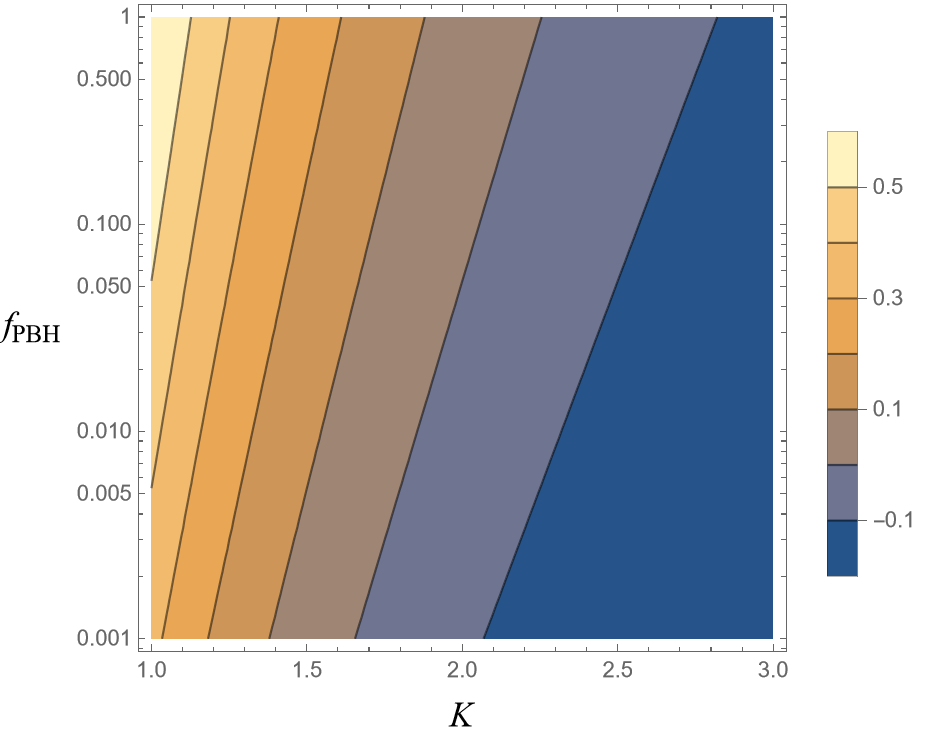}
	\caption{The $x$-axis is Sagan's parameter of the Kardashev civilization type. The $y$-axis is the fraction of PBH for a mass of $M_{\rm{\tiny{PBH}}} = 1 M_{\odot}$. The color bar of the contours shows the quantity $K_{\rm{achieved}}/K -1$, in which $K_{\rm{achieved}}$ is the amount of civilization parameter achieved from PBHs.} \label{fig:1}
\end{figure}

\begin{figure}
	\centering
	\includegraphics[scale=0.55]{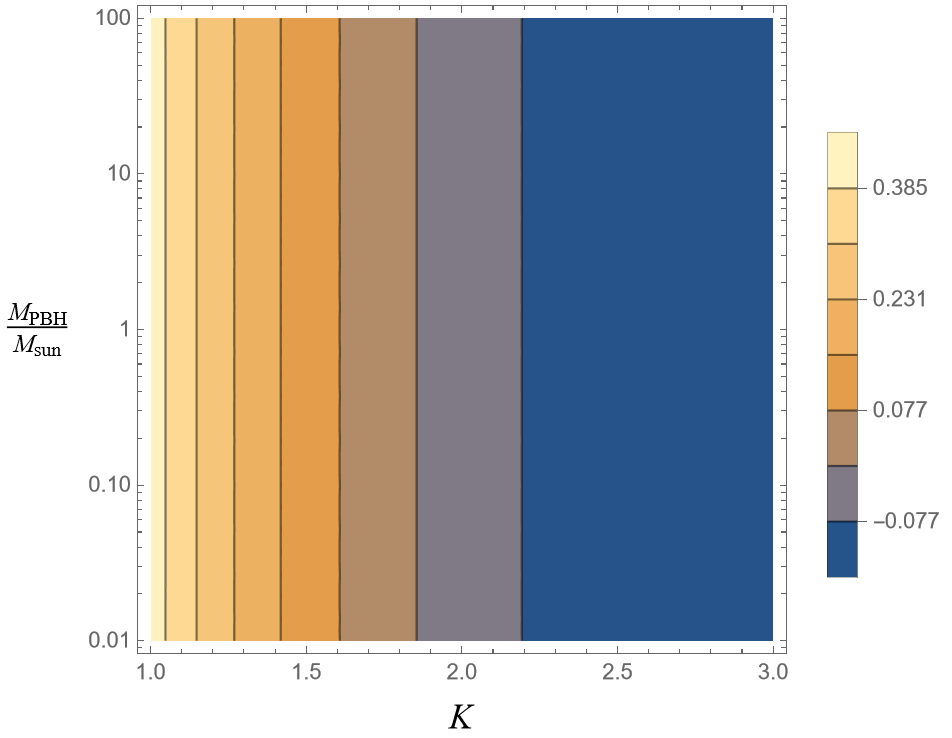}
	\caption{The $x$-axis is the Sagan's parameter of the Kardashev civilization Type $K$. The $y$-axis is the mass of PBH in solar mass for a fraction $f_{\rm{\tiny{PBH}}} = 0.01$. The colour-bar of the contours shows the quantity $K_{\rm{achieved}}/K -1$, which $K_{\rm{achieved}}$ is the amount of civilization parameter achieved by PBHs.} \label{fig:2}
\end{figure}

{{In figure \ref{fig:1}, we plot a contour representation of the chance that an advanced civilization will provide its whole energy from our proposal. It means that for a certain type of Sagan's parameter $K$, the energy is provided only by the Dyson sphere\textendash like structures around PBHs. The $x$-axis is Sagan's parameter of the Kardashev civilization type, which determines the energy needed for a civilization. The $y$-axis is the fraction of PBH for a mass of $M_{\rm{\tiny{PBH}}} = 1 M_{\odot}$. The color bar of the contours shows the quantity $K_{\rm{achieved}}/K -1$, in which $K_{\rm{achieved}}$ is the amount of civilization parameter achieved from our proposed method using only PBHs. The bluish regions (with negative values) are not suitable for our proposal. Note that the contour plot is obtained by following steps: First, by knowing the Sagan's parameter $K$ and equation \ref{eq:sed}, we can find the space exploration length. In the next step, by using equations \ref{eq:NPBH} and \ref{eq:Fcal}, we can find the number of PBH from the host planet of ET-AI up to the scale of ${\cal{D}}_{\rm{sed}}$. Now, the case, $K \leq K_{\rm{achieved}}$ means all energy can be obtained from PBHs. Figure  \ref{fig:1} shows that for a specific $K$ value, by increasing the fraction of PBH, $f_{\rm{PBH}}$, more energy can be provided (corresponding to $K_{\rm{achieved}}$). For a specific fixed $f_{\rm{PBH}}$, the chance of providing all of the energy from PBHs decreases at larger Sagan scales. This is a consequence of the fact that the amount of needed energy is higher than the energy which ET-AI can obtain from PBHs at larger distances. \\
		
In figure \ref{fig:2}, for a $f_{\rm{\tiny{PBH}}}=0.01$, we plot the contours of $K_{\rm{achieved}}/K -1$ for PBHs with mass in the range of $0.01-100 M_{\odot}$ as a function of Sagan's $K$ parameter. Again, we should note that $K_{\rm{achieved}}$ is obtained by the method we describe for figure \ref{fig:1}. The plot indicates that for a specific $K$ (space exploration length), the amount of achievable energy from our method is almost independent of the mass of the PBH. We anticipate this as the amount of energy from a PBH accretion disk is proportional to its mass. On the other hand, the abundance of smaller PBHs compromises the amount of energy each can deliver to a Dyson sphere\textendash like structure. The two plots show that the main parameter is the fraction of PBHs.}}

The figures show that there is a reasonable chance that in a proper mass range of PBHs and a nonzero fraction of PBHs serving as dark matter, an AI civilization uses PBHs as a source of energy. Note that we choose a conservative value for $\eta_{\rm{disk}}$. \\
{{To quantify this we provide a numerical expression for the minimum fraction of PBH ($f_{\rm{PBH|min}}$) to satisfy the condition  $K_{\rm{achieved}}/K -1 >0 $, (the boundary between the positive and negative regions of contour plot in figure \ref{fig:1} ), as a function of $K$.
		
		\begin{equation}
			f_{\rm{PBH|min}} \simeq 10^{5.0K -11.4}.
		\end{equation}
		In terms of space exploration length, this relation will be
		\begin{equation}
			f_{\rm{PBH|min}} \simeq 0.1\times \log(\frac{{\cal{D}}_{\rm{sed}}}{1\rm{ pc}}).
		\end{equation}
	Note that we normalize the space exploration length to $1\rm{pc}$.
}} Now, the question is about the observational consequences of the SET-AI.
The previous ideas are based on the technique for searching for Dyson spheres around stars in the Milky Way.
The idea is to look at the infrared excess radiation of stars. A specific example of this kind of study appeared in works of \cite{2022MNRAS.512.2988S,2024AJ....168..157C}. The Infrared excess can be an effect of circumstellar dust, protoplanetary disks, or debris disks, and it has a degeneracy with a Dyson sphere\textendash like structure. \\
For our proposal, we suggest that ET-AI uses PBHs as an energy source. Observing an excess in the submillimeter and infrared radiation in the interstellar medium could indicate a Dyson sphere. For an advanced civilization which brings a Dyson sphere to a temperature of $10 ~\rm{K}$ with limited efficiency, the waste energy will be in a blackbody radiation at $\sim 0.3 ~\rm{mm}$ from a structure with an extension of the order of $\sim 2700 ~ \rm{AU}$ around a PBH accretion disk. We obtain this scale by equation \ref{eq:Rdyson-pbh}, for a specific case where we assume a PBH mass of $1M_{\odot}$ and an efficiency of $\eta_{\rm{disk}}=10^{-4}$.\\
This submillimetre excess can be observed by the Atacama Large Millimeter/submillimeter Array (ALMA) by its Band 10 detector working at $ \sim 950~ \rm{GHz}$, as designed by the National Astronomical Observatory of Japan \citep{2013PhyC..494..189U}.
 {{This means that this excess can resemble a multiblackbody spectrum akin to the accretion disk of a PBH (see figure 3 in \cite{Hsiao:2021qij}). Their results are based on a multicolour blackbody radiation calculation of a disk with a Dyson sphere (see \cite{1984PASJ...36..741M,1986ApJ...308..635M, 2000MNRAS.313..193M}). However, in our scenario, the two peaks of the blackbody are separated further in wavelength. For the accretion disk, one should observe the spectrum in X-ray and UV, while for the Dyson sphere, one should go to submillimetre scales. On the other hand, the spatial resolution of ALMA in Band 10 is $\sim 0.5''$. This means a resolution of $\sim 2.42 \times 10^{-6} \rm{rad}$. Accordingly, for a  $\sim 2700 \rm{AU}$ extended space structure, for direct imaging, we can explore our Galaxy up to scales $\sim 5.4 ~\rm{kpc}$. However, this technique needs a concrete study closely tied to ALMA's sensitivity, depth, and required exposure time for a detection, which is out of the scope of the current work. To distinguish these images from other celestial objects, one has to get spectroscopy of the Dyson sphere candidate as well.}} For a solar\textendash mass PBH with $\eta_{\rm{disk}}\sim 10^{-4}$ and efficiency of $30\%$ for the Dyson sphere,  we will have solar Luminosity  $ \sim 1 L_{\odot}$ at submillimeter in a blackbody radiation with angular resolution of $(\frac{2700 \rm{AU}}{ x_{\rm{PBH}}}) \rm{rad}$, where $x_{\rm{PBH}}$ is the distance from the PBH in astronomical units. This excess in blackbody shape must be compared with the same wavelength radiation in the interstellar medium for detection. Also, we anticipate an excess at higher frequencies due to the accretion disk at that region. In other words, the Dyson sphere\textendash like structure radiation is a signal we want to extract from measurement of interstellar radiation while accounting for noise at the same wavelength. \\
Another observational consequence of our work is the quest for the SET-AI at the extragalactic scales.
This means that the galaxies will be in a state of excess radiation in the infrared and submillimeter for the scales proportional to SED. Accordingly, one target of the future SET-AI will be galaxies with considerable infrared radiation. For a specific example let us consider the following. Assume an advanced civilization with Sagan's scale $K \simeq 2.2$, with the assumption of a fraction of the $f_{\rm{PBH}}=1.0$, in which the idea of using PBHs as a source of energy is achievable. The space exploration length in this configuration is in order of $\sim 4.8{\rm{pc}}$. It means that with ALMA Band 10 with a given resolution of $\sim 0.5''$, we can probe the nearby galaxies up to $\sim 2~ {\rm{Mpc}}$ for  excess at $\sim 0.3 ~\rm{mm}$ is spectra and direct imaging.
The wavelength mentioned again is suitable for Dyson sphere\textendash like structures with $\sim 10 \rm{K}$.
For galaxies at cosmological distances, one can refer to an interesting way of quantifying the detection of the infrared excess: looking for excess infrared radiation in a plot versus the UV continuum slope. For more details see \cite{2016ApJ...833...72B}. In distant galaxies, the angular resolution is not enough for AI-designed structure detection. However,  one can find a sign of ET-AI for further investigation by excess radiation in higher redshifts.  
\section{Conclusions and Remarks} \label{Sec4}
{{The search for life in our Universe is a vast arena of study. In this work, we propose the idea of SET-AI as an advanced civilization using the proposal that ET-AI uses interstellar Dyson sphere\textendash like megastructures to harvest energy. For the measure of civilization advancement, we introduce the space exploration length, or SED. We propose a relation between this length and the Kardashev scale.
One step further, we argue that PBHs as plausible candidates of DM, are distributed evenly in the DM halo. They are more abundant and efficient energy sources than the main\textendash sequence stars for reaching and harvesting energy beyond the star\textendash planet system. We investigate the percentage of Kardashev scale achievement as a function of the fraction of PBHs as DM and the masses of PBHs, as shown in Figures \ref{fig:1} and \ref{fig:2}.\\ 
		We explore the prospects of detecting ET-AI. On a galactic scales, one can search for infrared and submillimeter excess in the interstellar medium plus the feature of a multiblackbody spectrum. ALMA is a suitable telescope for searching for these structures, even for direct imaging up to scales of $\sim 5.4 \rm{kpc}$. These investigations apply to the galaxies beyond our Milky Way, and we can consider the modified Drake equation. Beyond the Milky Way with the ALMA telescope, one can search for a Dyson sphere\textendash like structure up to $2 \rm{Mpc}$ and for higher redshifts, one can look at the excess infrared radiation in a plot versus the UV continuum slope of a galaxy.
		In future studies in this direction, we will focus on the dependency of the accretion efficiency of PBHs on their mass, redshift, and environment. The galaxy properties can also affect the results of the achieved Kardashev parameter. The observational hints of this idea need further and extensive investigation.
		In conclusion, it seems that it is worth paying attention to finding ET-AI when we are looking for life in the cosmos.}}



\section*{ACKNOWLEDGMENTS}
This work is dedicated to Sohrab Rahvar, whom we admire for first teaching us about life in the Universe. \\
We thank the anonymous referee for carefully reading our manuscript and for their insightful comments and suggestions, which improved the manuscript extensively. \\  
S.B. is partially supported by the Abdus Salam International Center for Theoretical Physics (ICTP) under the regular associateship scheme.
S.B. is partially supported by the Sharif University of Technology Office of Vice President for Research under grant No. G4010204.

\iftrue

\fi

\end{document}